\documentclass[pss]{wiley2sp} 
\usepackage{amsmath,amssymb}

\begin{document}

\title{Sample Size Effects on the Transport Characteristics of Mesoscopic Graphite Samples}

\titlerunning{Sample Size Effects}

\author{%
  J. Barzola-Quiquia\textsuperscript{\textsf{\bfseries }},
J.-L. Yao\textsuperscript{\textsf{\bfseries }}, P.
R\"odiger\textsuperscript{\textsf{\bfseries }}, K.
Schindler\textsuperscript{\textsf{\bfseries }}, P.
Esquinazi\textsuperscript{\textsf{\bfseries \Ast}}}

\authorrunning{Barzola-Quiquia et al.}

\mail{e-mail
  \textsf{esquin@physik.uni-leipzig.de}, Phone
  +49-341-9732751, Fax +49-341-9732769}

\institute{%
  \textsuperscript{~}\,Division of Superconductivity and Magnetism,
Universit\"{a}t Leipzig, Linn\'{e}stra{\ss}e 5, D-04103 Leipzig,
Germany}

\received{XXXX, revised XXXX, accepted XXXX} 
\published{XXXX} 


\abstract{%
%
%
%
\abstcol{In this work we investigated correlations between the
internal microstructure and sample size (lateral as well as
thickness) of mesoscopic, tens of nanometer thick graphite
(multigraphene) samples and the temperature $(T)$ and field $(B)$
dependence of their electrical resistivity $\rho(T,B)$. Low energy
transmission electron microscopy reveals that the original highly
oriented pyrolytic graphite material -- from which the multigraphene
samples were obtained by exfoliation -- is composed of a stack of
$\sim 50~$nm thick and micrometer long crystalline regions separated
by interfaces running parallel to the graphene planes. We found a
qualitative and quantitative change in the behavior of $\rho(T,B)$
upon thickness of the multigraphene samples, indicating that their
internal microstructure is important.} {The overall results indicate
that the metallic-like behavior of $\rho(T)$ at zero field measured
for bulk graphite samples is not intrinsic of ideal graphite. The
results suggest that the interfaces between crystalline regions may
be responsible for the superconducting-like properties observed in
graphite. Our transport measurements also show that reducing the
sample lateral size as well as the length between voltage electrodes
decreases the magnetoresistance, in agreement with recently published
results. The magnetoresistance of the multigraphene samples shows a
scaling of the form ($(R(B) - R(0))/R(0))/T^\alpha = f(B/T)$ with a
sample dependent exponent $\alpha \sim 1$, which applies in the whole
temperature 2~K~$\le T \le 270~$K and magnetic field range $B \le
8~$T.}}


\maketitle   

\section{Introduction}

Ideal graphite consists on layers of honeycomb lattices of carbon
atoms, characterized by two non-equivalent sites, A and B, in Bernal
stacking configuration (ABABAB$\ldots$). Although there is not yet
consent on the interlayer cohesive or binding energy per carbon atom
between graphene layers in ideal graphite \cite{kelly}, several
experimental facts suggest that this coupling is not larger than
$\sim 0.05~$eV~\cite{yakadv07}. The huge anisotropy in the
resistivity (ratio of the in-plane divided out-of-plane
resistivities) $10^6 \sim \rho_a/\rho_c \gtrsim 10^4$ at low and room
temperatures, respectively,  in good quality samples  indicate the
quasi two-dimensionality of the transport in the graphene layers of
the graphite structure~\cite{yakadv07}. On the other hand, the
effects of the real microstructure, the size and quality of the
crystalline regions and their interfaces or even the influence of the
overall sample size on the transport properties in single graphene,
tens of graphene layers (multigraphene) as well as in bulk graphite
samples are still not well studied.

A quick comparison between the published temperature ($T$) dependence
of the resistivity $\rho(T)$ at zero magnetic field ($B$)  suggests
that $\rho(T)$ behaves differently in graphene and in high quality
graphite. Whereas highly oriented pyrolytic graphite (HOPG) with
narrow rocking curve widths shows a $\rho(T)$ that decreases
decreasing $T \lesssim 200~$K, $\rho(T)$ of single graphene layers
appears to steadily increase decreasing $T$, keeping the electron
density $n$ low enough (see for example \cite{tan07} and references
inside). Although the influence of the substrates on the electronic
transport of graphene is significant and cannot be neglected
\cite{chen08}, the overall published experimental  data do suggest
that the ubiquitous decrease of $\rho(T)$ (at $T \lesssim 200~$K)
appears only in good quality HOPG samples \cite{yakovadv03}. It is
interesting to note that it remains still unclear, which is the
absolute resistivity (parallel and perpendicular to the graphene
planes) as well as the temperature dependence of defect free, ideal
graphite. This is a basic question that still remains fully open due
to the sensitivity of the graphite structure to defects. The results
presented in these studies suggest that the resistivity of ideal
graphite is not metal-like, i.e. $\rho(T)$ decreases with $T$ in all
the $T-$range.

Our work provides experimental hints on the correlations between the
internal structure of the samples and their transport properties. The
results are not only important to understand ideal graphite/graphene
properties but also to understand the role of defects and/or
interfaces within the graphite structure. Recently done high
resolution magnetoresistance measurements on multigraphene samples
show anomalous hysteresis loops below a ``critical" temperature that
indicates the existence of superconducting grains with high critical
temperature embedded in a semiconducting matrix \cite{esqui08}. Those
results are the last ones of a series of experimental hints
suggesting the existence of granular superconductivity in graphite
(for a short review see Ref.~\cite{kopejltp07}). The data presented
in this  study provide a hint where these superconducting regions
might be located.

\paragraph{Size effects}

Recently published experimental work on highly oriented pyrolytic
graphite (HOPG) showed that the change of the electrical resistance
with magnetic field, i.e. the ordinary magnetoresistance (MR),
decreases with the sample size even for samples hundreds of
micrometer large \cite{gon07}. This effect was ascribed to the large
carrier mean free path $\ell$ as well as the large Fermi (or
de~Broglie) wavelength $\lambda_F$ in graphite. Recently, Garc\'ia et
al. \cite{gar08} reported the development of  an experimental method
and its theoretical basis to obtain without free parameters these two
transport properties based on the measurement of the resistance
through micro-constrictions on a $\sim 30~\mu$m thick HOPG sample. In
that work micrometer large values for $\ell$ and $\lambda_F$ at $T
\le 10~$K were obtained in agreement with the expectations from the
magnetoresistance results\cite{gon07}. The carriers with the largest
mean free path have $\ell \sim 10~\mu$m appear to be limited by the
crystallite size in the used high quality HOPG sample \cite{gar08}.

In the last 50 years, the transport properties of graphite have been
interpreted in terms of two- (three-) band Boltzmann-Drude
approach\cite{kelly}. However, the use of the standard approaches to
understand the electrical properties of graphite is doubtful since
due to the large carrier mean free path  ballistic instead of
diffusive transport should be applied.  The large Fermi wavelength
(due to the small carrier density of the carriers $\lambda_F \sim
1~\mu$m) implies that diffraction effects within the sample may play
also a role. Moreover, the differences in the transport properties
between apparently similar samples together with  electron force
microscopy (EFM) results obtained on HOPG samples \cite{lu06} provide
further evidence that HOPG samples should be considered as a
non-uniform electronic system. This non-uniformity is not an
intrinsic property but it depends on parameters like the defect
density or interfaces within the measured sample region and therefore
we expect that the electrical resistance will depend on the measured
sample size. The overall results presented in this study especially
the sample size effects confirm once more the inadequacy of the
standard models to understand the transport properties of graphite.
All these results as well as the possibility that granular
superconductivity could influence the transport of HOPG cast doubts
on the applicability of the theoretical descriptions as has been done
in the literature up to date to understand the transport properties
of graphite.

\paragraph{Single graphene, multigraphene and the role of the microstructure}

Nowadays there is a general interest of the solid state community on
the transport properties of single graphite layers dubbed graphene
\cite{kat07}. As in the very first transport experiments done in
graphene \cite{novo05,zhang05}, these samples are usually fixed on
substrates. In general neither the possible variations of the
electrical potential at the surface of the substrates as well as
their shape variations nor the influence of the environment were
considered important issues that may influence the transport
properties of the graphene samples. Recent experimental evidence
obtained in suspended graphene samples appears to confirm the
detrimental effect of the substrates on the mobility of graphene
\cite{bol08,du08}. The lowest achievable electronic density ($n
\gtrsim 10^{10}~$cm$^{-2}$) in free standing as well as
fixed-on-substrate graphene samples is still far away from the Dirac
point, due to defects in the graphene structure, intrinsic bending or
due to the influence of the substrate itself. This restriction limits
 the Fermi wavelength, $\lambda_F$, and therefore  the largest
achievable mobility ($\mu = (e/h)\ell \lambda_F$) since the carrier
mean free path $\ell$ cannot be larger than the size of the graphene
samples. In fact, recently published work \cite{chen08} showed that
the minimum conductivity is governed not by the physics of the Dirac
point singularity but rather by carrier-density inhomogeneities
induced by the potential of charged impurities that may come from the
substrate.

One possibility to overcome these limitations is to use several
nanometers thick multigraphene samples, which are much less sensitive
to the environment as well as to the substrates. In high quality
graphite samples the graphene layers inside are of larger perfection
than single graphene layers. Therefore it should be possible to
obtain their intrinsic transport properties without external
influence other than those coming from the lattice defects. A direct
proof of the perfection of the graphene layers in high quality
graphite is given by the very low carrier density measured at low
temperatures, which is nearly two orders of magnitude smaller than
the minimum obtained for suspended graphene samples \cite{gar08}.

\paragraph{In this work} we
studied the temperature and magnetic field response of the electrical
resistivity of multigraphene samples with thickness between 10~nm and
20$~\mu$m. Further characterization of the microstructure was done
with a low voltage transmission electron microscope (TEM) that
allowed us to observe some details of the internal structure of the
samples as the typical distance between the interphases separating
crystalline regions. Micro-Raman studies on some of the nanometer
thick samples were also performed. We provide evidence for two size
effects on the transport properties of multigraphene samples. One is
related to the thickness and correlates to the internal
microstructure of the samples. The second size effect deals with the
influence of the sample lateral size on the transport properties. The
paper has four more sections. Section~\ref{details} explains some
details of the experimental methods we used. Section~\ref{charac}
describes the measured samples and their characteristics and includes
the TEM and Raman results. Section~\ref{tra} shows the main transport
results. This section is divided in three subsections where we
describe the different behavior of the transport properties upon the
sample size. The main conclusions are given in Sec.~\ref{con}

\section{Experimental Details} \label{details}

\paragraph{Sample preparation}
In order to carry out a systematic study we have performed
measurements in different tens of nanometer thick multigraphene
samples obtained by exfoliation from the same highly oriented
pyrolytic sample with a mosaicity of $0.4^\circ \pm 0.1^\circ$
(HOPG(0.4)).  One part of the original HOPG sample was left with a
thickness of $\sim 20~\mu$m, which transport properties resemble the
usual behavior observed in high-quality HOPG of similar
characteristics \cite{yakovadv03}.

The initial HOPG material of dimensions $ 4 \times 4 \times
0.5$~mm$^3$ was glued on a substrate using GE~7031 varnish. We used a
simple technique to produce the multigraphene films, which consists
in a very carefully mechanical press and rubbing the initial material
on a previously cleaned substrate. As substrate we used p-doped Si
with a 150~nm SiN layer on top. This substrate helps to select the
multigraphene films because -- in comparison with Si substrates with
a top layer of SiO$_2$ -- the SiN layer provides a higher color
contrast allowing us to use optical microscopy to select the film.
After the rubbing process we put three times  the substrate
containing the multigraphene films in a ultrasonic bath during 2~min
using high concentrate acetone. This process cleans and helps to
select only the good adhered multigraphene films on the substrate.
After this process we used  optical microscopy and later scanning
electron microscopy (SEM) to select and mark the position of the
films. For the production of the electrical contacts we used
conventional electron lithography process. Afterwards   the contacts
were done by thermal deposition of  Pd (99,95\%) in high vacuum
conditions.  We have used Pd because it does not show any Schottky
barrier when used with carbon. Measurements of the resistance of the
Pd-electrodes alone showed negligible magnetoresistance. For the
transport measurements the sample was glued on a chip carrier. The
contacts from the chip carrier to the electrodes on the sample
substrate were done using a $25~\mu$m gold wire fixed with silver
paste.

The advantage of using HOPG of good quality is that in these samples
and due to the perfection of the graphene layers and low coupling
between them, a low two-dimensional carrier density $2 \times
10^8~$cm$^{-2} \lesssim n \lesssim 10^{11}~$cm$^{-2}$ in the
temperature range 10~K~$ \lesssim T \lesssim 300~$K is obtained
\cite{gar08}. The carrier density values obtained in
Ref.~\cite{gar08} are smaller than in typical few layer graphene
(FLG) samples probably due to lattice defects generated by the used
method to produce them and/or surface doping \cite{bol08,du08}.

\paragraph{Transport measurements} Low-noise four-wires (two for the input
currents and two for the voltage measurement) resistance measurements
have been performed by AC technique (Linear Research LR-700 Bridge
with 8 channels LR-720 multiplexer) with ppm resolution and in some
cases also with a DC technique (Keithley 2182 with 2001 Nanovoltmeter
and Keithley 6221 current source). The temperature stability achieved
was $\sim 0.1~$mK and the magnetic field, always applied normal to
the graphene planes, was measured by a Hall sensor -- just before and
after measuring the resistance -- and located at the same sample
holder inside a superconducting-coil magneto-cryostat. We used
current amplitudes between $1~ \textrm{and}~ 100~\mu$A.  The
measurement of the resistance at different positions of the same
sample  indicate that the contact resistance contributions in the
absolute value as well as in the temperature and magnetic field
measurements are negligible, as expected due to the four-wire
configuration used. The magnetoresistance measurements were done
always with the magnetic field applied perpendicular to the graphene
planes, i.e. parallel to the samples c-axis.

\paragraph{Transmission Electron Microscopy}
The images of the internal structure of the HOPG sample were obtained
 using a Nova NanoLab~ dual beam microscope from the FEI company (Eindhoven).
 A HOPG lamellae was prepared for transmission electron microscopy
(TEM) using the in-situ lift out method of the microscope. The TEM
lamellae of HOPG was cut perpendicular to the graphene layers.
Therefore, the electron diffraction provided information on the
crystalline regions and their defective parts parallel to the
graphene layers. After final thinning, the sample was left on a TEM
grid. A solid-state scanning transmission electron microscopy (STEM)
detector for high-resolution analysis of thinned samples was used.
The voltage applied to the electron column was 18 kV and the currents
used were between 38 to 140 pA.

\paragraph{Micro-Raman Spectroscopy}
Raman spectra of multigraphene samples were  obtained at room
temperature and ambient pressure with a Dilor~XY~800 spectrometer at
514.53~nm wavelength (Green) and a $2~\mu$m spot diameter. The
incident power was varied between 0.5 to 3~mW to check for possible
sample damage or laser induced heating effects. No damage and
significant spectral change was observed in this range of incident
power.

\section{Samples Characteristics} \label{charac}

\begin{figure*}[htb]%
\includegraphics*[width=\textwidth,height=12cm]{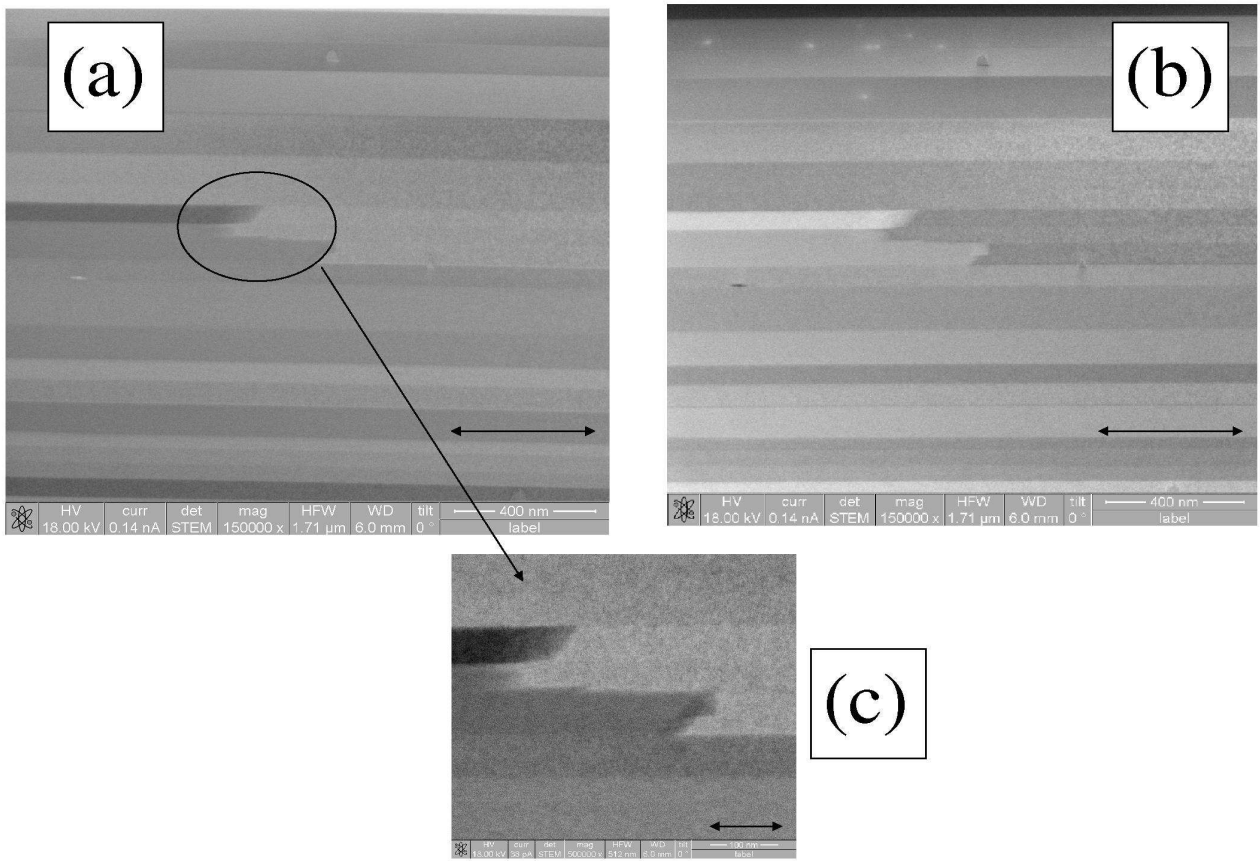}
\caption{Transmission Electron Microscopy pictures taken parallel to
the graphene layers of the HOPG lamelle. The c-axis is perpendicular
to the clearly observable stripes of different gray colors, each
representing a crystalline region with a slightly different
orientation. The arrows in (a) and (b) indicate 400~nm length scale
and in (c) 100~nm.} \label{TEM}
\end{figure*}
 Table~\ref{table} shows the sample dimensions and
names. The error bars in the thickness are the maximum estimated
ones, taking into account the maximum error in the measurement and
calibration of the optical and/or atomic force microscope (AFM) as
well as the irregularities in the samples borders. The error in the
absolute value of the resistivity takes those errors into account as
well as errors in the width and length. Optical pictures of some of
the measured samples are included in the figures below.

\begin{table}[b]
  \caption{Samples' names, resistivity and size.}
  \begin{tabular}[htbp]{@{}lllll@{}}
    \hline
    Name & resistivity  & thickness & length & width \\
    &$\rho(4$K$) [\mu\Omega$cm]&&&\\
    \hline
    HOPG & $ 6 \pm 2$ & $17 \pm 2~\mu$m & 4.4~mm & 1.1~mm\\
    L5 & $70 \pm 15$ & $12 \pm 3$~nm  & $27~\mu$m & $14~\mu$m\\
    L8A & $480 \pm 100$ & $13 \pm 2$~nm & $14~\mu$m & $10~\mu$m\\
    L2A & $130 \pm 33$ & $20 \pm 5$~nm & $5~\mu$m & $10~\mu$m\\
    L8B & $124 \pm 25$ & $45 \pm 5$~nm & $3~\mu$m & $3~\mu$m\\
    L7 & $29 \pm 6$ & $75 \pm 5$~nm & $17~\mu$m & $17~\mu$m\\
    \hline
  \end{tabular}
  \label{table}
\end{table}

\subsection{Transmission Electron Microscopy Results}\label{tem}
Figure~\ref{TEM} shows the bright field (a) and dark field (b)
details obtained with the low-voltage STEM. Figure~\ref{TEM}(c) shows
a blow out of a detail of (a). The different gray colors indicate
crystalline regions with slightly different orientations.
 The images indicate that the average thickness of the crystalline
 regions is $60 \pm 20~$nm. One can also resolve the interfaces
 perpendicular to the c-axis of the layers and  between the regions
 as well as the end parts of the crystalline regions along the graphene layers direction,
 see Fig.~\ref{TEM}(c). Electron back scattering diffraction
measurements done on similar HOPG samples indicate that the typical
size of the single crystalline regions (on the (a,b) plane) ranges
between 1 to $\sim 10~\mu$m \cite{gar08}. If the interface between
the crystalline regions as well as the defects in the crystalline
lattice have some influence on the transport properties we would
expect to see a change in the behavior of the transport properties
between samples of thickness of the order or less than the average
thickness of the crystalline regions.

\subsection{Raman Results}\label{Raman}
The Raman spectrum of graphite \cite{reich04}, a few layers and
single graphene \cite{graf06,fer06} have been thoroughly measured and
discussed in recent published studies and review. In this study we
are interested in three  Raman maxima, namely the G-peak around
1582~cm$^{-1}$ due to a Raman active in-plane optical phonon
E$_{2g}$, its neighbor peak, the D-line around 1350~cm$^{-1}$, which
is very sensitive to the amount of lattice disorder, and the
D'-line\cite{graf06} (or 2D peak\cite{fer06}) at $\sim
2700~$cm$^{-1}$ and its splitting in two or more maxima upon the
number of graphene layers the sample has \cite{graf06}.

\begin{figure}[t]%
\includegraphics*[width=\linewidth,height=12.0cm]{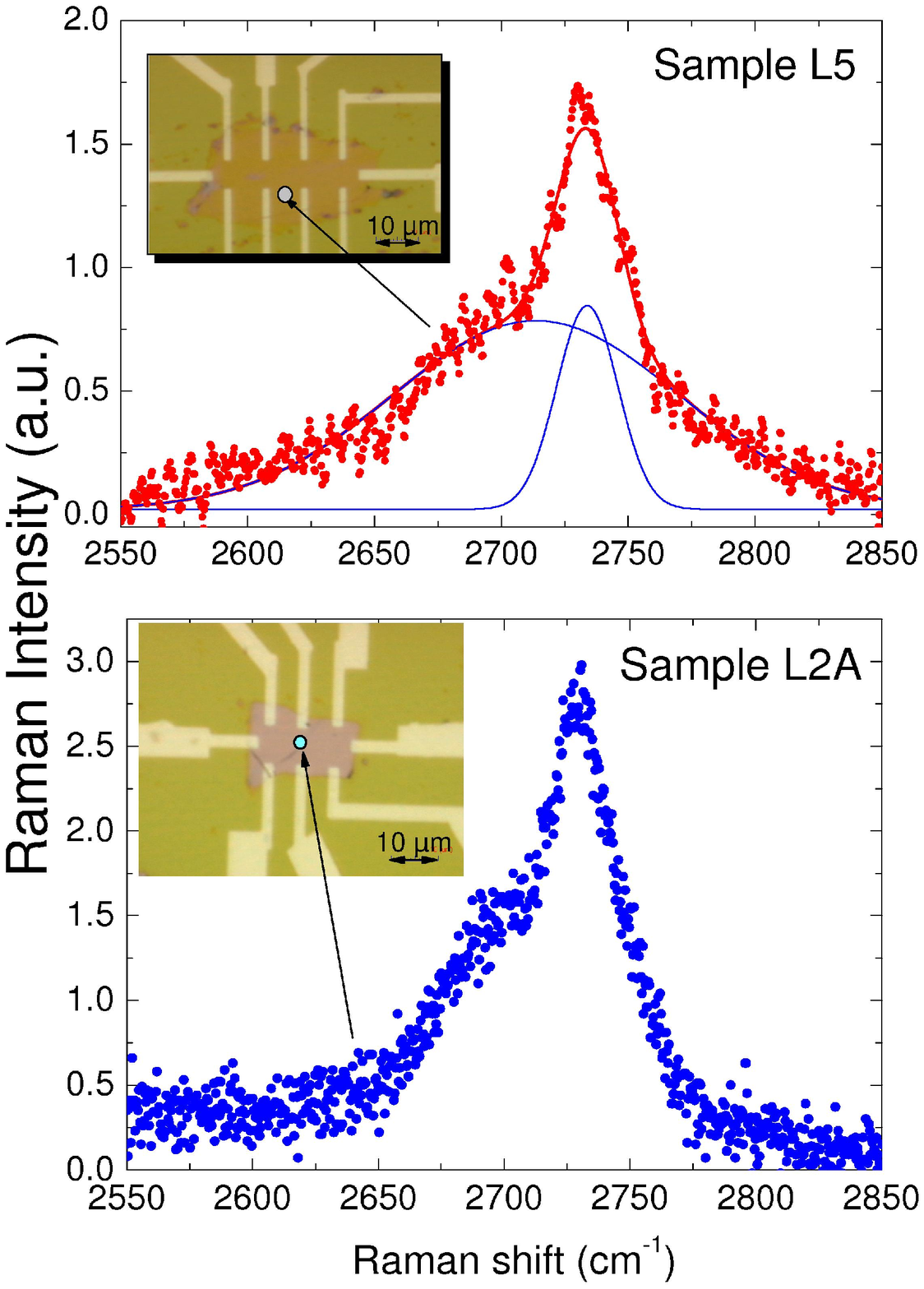}
\caption{Raman spectra around the D'-line for samples L5 (upper
picture) and L2A (bottom picture). The continuous lines in the upper
picture represent the Lorentzian peaks used to fit the data. The
insets show optical microscope pictures of the samples with the array
of Pd electrodes.} \label{raman2700}
\end{figure}

Figure~\ref{raman2700} shows the Raman spectra of samples L5 (upper
picture) and L2A (bottom) around the D'-line. The splitting into two
peaks of the D'-line is in agreement with recently published results
\cite{graf06}. The main central narrow peak is at $2732 \pm
2~$cm$^{-1}$ whereas the value of the splitting is $21 \pm
2~$cm$^{-1}$ in very good agreement with that observed for HOPG bulk.
Taking into account that both samples L5 and L2A have a thickness
between 10 and 20~nm, this difference in thickness does not affect
the Raman D'-line. Taking this line as reference one would conclude
that both samples are identical.

\begin{figure}[b]%
\includegraphics*[width=\linewidth,height=10cm]{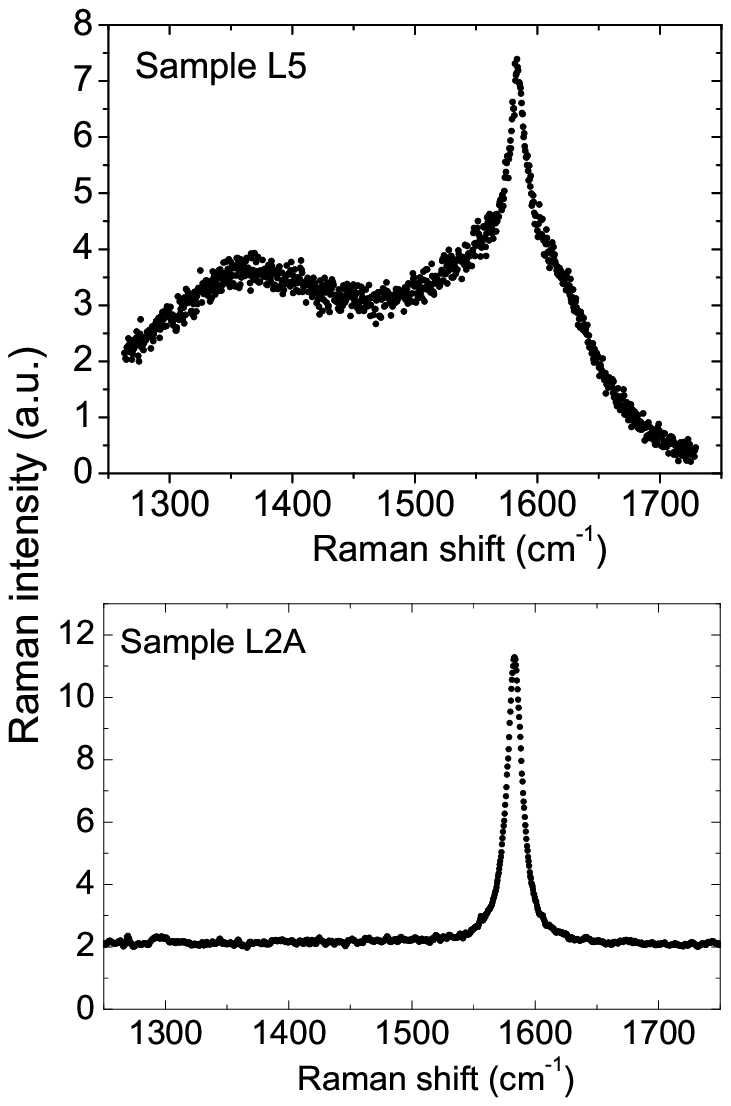}
\caption{Raman spectra around the D- and G-line (1350 cm$^{-1}$ and
1582 cm$^{-1}$)) for samples L5 (upper picture) and L2A (bottom
picture).} \label{raman1350}
\end{figure}

However, according to literature the structural quality of the
samples can be resolved by investigating the D-line at $\sim
1350~$cm$^{-1}$. We note that also the edges of the samples as well
as the borderlines between regions of different thicknesses may
contribute to the D-band signal. The Raman spectra between  1300 and
1700 cm$^{-1}$ have been measured at different positions of sample L5
and in sample L2A. The results are shown in Fig.~\ref{raman1350} for
the two samples. The broad D-line at $\sim 1350~$cm$^{-1}$ is clearly
seen in sample L5 (at the same position as in Fig.~\ref{raman2700})
but it is completely absent in sample L2A. We checked that a similar
curve is observed at different positions of the sample L5. The small
peak at $\sim 1300~$cm$^{-1}$ in sample L2A is due to the substrate.
The G line at $\simeq 1582~$cm$^{-1}$ is observed for both samples,
see Fig.~\ref{raman1350}. From these results we would conclude that
sample L5 has more disorder or that the borders or edges have a
larger influence to the Raman spectra than in sample L2A. From
literature\cite{peres06} we would expect that this disorder should
have some influence on the transport properties of graphene as well
as in a multigraphene sample.

\section{Transport Measurements}\label{tra}

\subsection{Thickness and Temperature dependence of the
resistivity at zero magnetic field}\label{ttd}

\begin{figure}[htb]%
\includegraphics*[width=\linewidth,height=7cm]{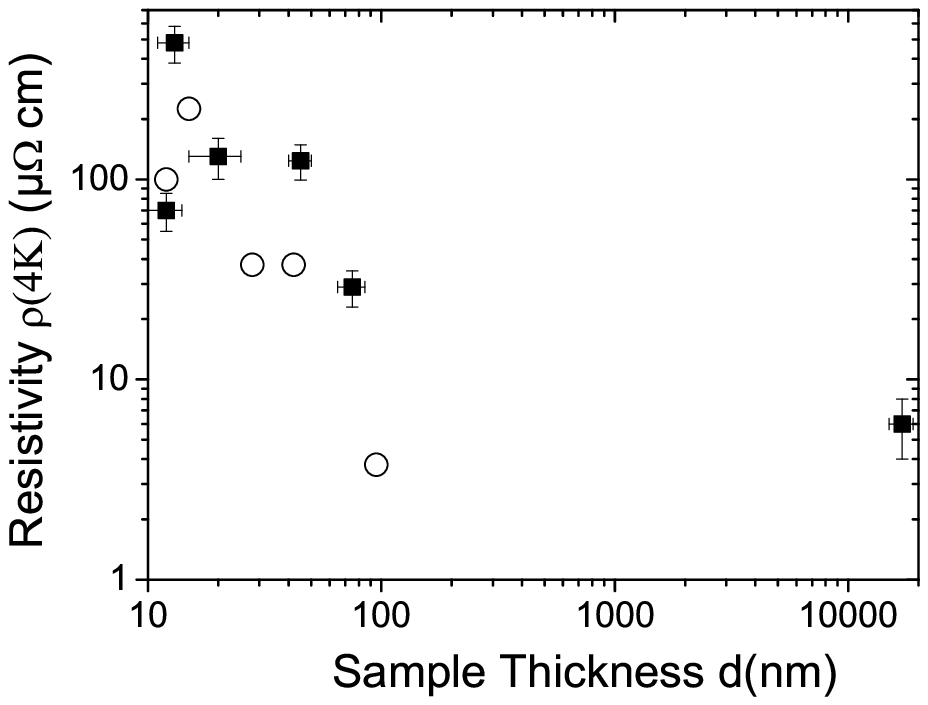}
\caption{Absolute resistivity at 4~K vs. sample thickness $d$ for the
six measured samples of this work ($\blacksquare$). The data points
($\bigcirc$) are taken from Ref.~\cite{kim05} as explained in the
text.} \label{rd}
\end{figure}

Figure~\ref{rd} shows the resistivity at 4~K of the six measured
samples vs. their thickness. It is clearly  seen that the resistivity
decreases increasing the sample thickness. The average change in
resistivity between $\sim 10~$nm
 to $17~\mu$m thick samples is  about two orders of magnitude, far beyond
 geometrical errors. A similar behavior was observed recently in
 multigraphene samples obtained with a different, micromechanical
 method\cite{kim05}.
 Because in that work no explicit absolute
 values of the resistivity at 4~K were given, we estimate it taking the
in that work given mobility, assuming that the carrier density does
not depend on the thickness and fixing arbitrarily the value of
$100~\mu\Omega$cm for the 12~nm thick sample reported in
Ref.~\cite{kim05}. The open circles shown in Fig.~\ref{rd} are the
data points from Ref.~\cite{kim05}. A reasonable agreement between
the two independently obtained measurements is obtained that speaks
for the reproducibility of the observed dependence.

\begin{figure}[htb]%
\includegraphics*[width=\linewidth,height=7cm]{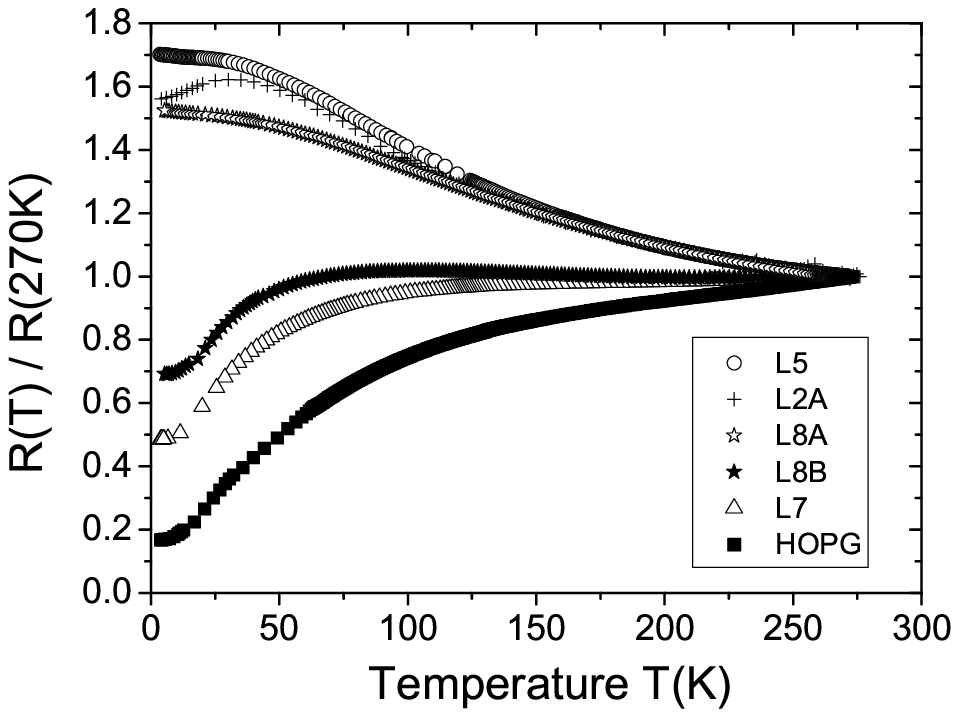}
\caption{Normalized resistance vs. temperature at zero applied field
for all the samples measured in this work.} \label{rt}
\end{figure}

The authors in Ref.~\cite{kim05} suggested that the decrease of
mobility $\mu$ (i.e. an increase in the resistivity at constant
carrier density) decreasing sample thickness provides an evidence for
boundary scattering. Taking into account the fact that one graphene
layer shows finite mobility \cite{bol08,du08}, boundary scattering is
certainly not the correct explanation for the observed behavior. A
possible explanation for the observed trend
 is that the larger the thickness the larger is the amount of defects
 and interfaces in the sample, see Fig.~\ref{TEM}, that produces the
 decrease in the resistivity. Since HOPG is a highly anisotropic
 material with huge anisotropy in the resistivity, it appears
 reasonable to assume that certain kind of lattice defects (vacancies, dislocations, etc.) may
 produce a sort of short circuits between layers, changing the dimensionality of
 the carrier transport and decreasing the resistivity.
 It appears unlikely, however, that randomly distributed
 point-like lattice defects can be the reason for the observed behavior. The results suggest
 the existence of  a
 kind of thickness threshold  around $\sim 10~$nm for the muligraphene samples
 obtained from HOPG(0.4) bulk graphite, see Fig.~\ref{rt}.

Regarding the two samples characterized
 with Raman we note that the sample L5, which shows a D-line (see
 Fig.~\ref{raman1350}) presumably due to the contribution of lattice disorder,
 has a lower resistivity than sample L2A that shows no D-line. In
 Ref.~\cite{peres06} was shown that extended defects in graphene can
 lead to self doping. Moreover, the presence of such defects can
 still lead to long carrier mean free path and a
 decrease in the resistivity. Our results speak
 for a non-simple influence of defects in graphite.

Figure~\ref{rt} shows the normalized resistivity as a function of
temperature  at zero magnetic field for all measured samples. The
overall results agree with those obtained for multigraphene samples
in Ref.~\cite{kim05}. However, we note the following, interesting
details:\\
- There is an apparent difference in the $T-$dependence between thick
and thinner samples. The samples thicker than $\sim 20~$nm show a
rather metallic behavior whereas thinner samples a semiconducting
like, see Fig.~\ref{rt}.\\
- There is basically no difference in the $T-$dependence between
sample L5 and L2A, with exception of the region at $T < 30~$K where
$R$ decreases decreasing $T$ in the thicker sample~L2A. This fact
indicates that the disorder the Raman D-line indicates does not
affect strongly the $T$-dependence of the resistivity.\\
- The overall results indicate that the maximum at $\sim 30~$K in the
resistivity observed in sample L2A may have the same origin as the
decrease in the resistivity below $\sim 100~$K observed  in thicker
samples, see Fig.~\ref{rt}. Taking into account the TEM results, see
Fig.~\ref{TEM}, we would conclude that this metallic-like behavior is
not intrinsic to ideal, defect-free graphite or multigraphene but it
is due to the influence of the interfaces inside the samples with
large enough thickness.

The obtained results suggest that the true $T-$dependence of the
resistivity in an ideal, defect-free multigraphene sample should be
semiconducting-like, i.e. it should increase decreasing temperature.
This dependence is actually the one expected for an ideal semimetal
with zero or small energy gap, in agreement with the large decrease
in carrier density decreasing temperature recently obtained in HOPG
samples \cite{gar08}.

\subsection{Magnetoresistance: Thickness dependence and scaling}\label{OMR}

\begin{figure}[htb]%
\includegraphics*[width=\linewidth,height=9.5cm]{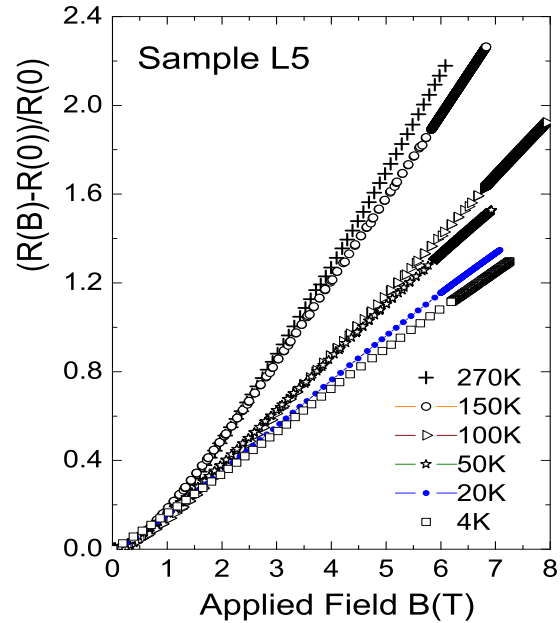}
\caption{The magnetoresistance  at different constant temperatures
for sample L5. The selected voltage electrodes were the two with the
largest distance, near the current input electrodes, see
Fig.~\protect\ref{raman2700}.} \label{omrl5}
\end{figure}

The MR was measured for a few samples. Here we concentrate ourselves
on two samples L5 and L7, see Fig.~\ref{rt}, which show a clear
difference in the $T-$dependence of the resistivity. The results for
these two samples are typical ones. The MR depends mostly on the
behavior of the resistivity, i.e. whether it shows a semiconducting
behavior like in samples L2, L2A (above 50~K) and L8A, or a metallic
one, like in samples HOPG, L8B and L7.

Figure~\ref{omrl5} shows the MR vs. magnetic field defined as MR$~=
(R(B)-R(0))/R(0)$ at different constant temperatures for sample L5.
Above a field of $\sim 1~$T the MR increases with temperature. At
lower fields it remains nearly $T-$independent. This increase of MR
with $T$, although anomalous,  it is actually what one expects if the
resistance decreases with $T$, as is the case for sample L5. Note,
however, that the MR is much weaker than the one measured in HOPG or
Kish graphite bulk samples where the MR is larger than 1000~\% at
fields above  0.5~T and at low temperatures
\cite{yakovadv03,heb05,tak04}.

\begin{figure}[htb]%
\includegraphics*[width=\linewidth,height=10.5cm]{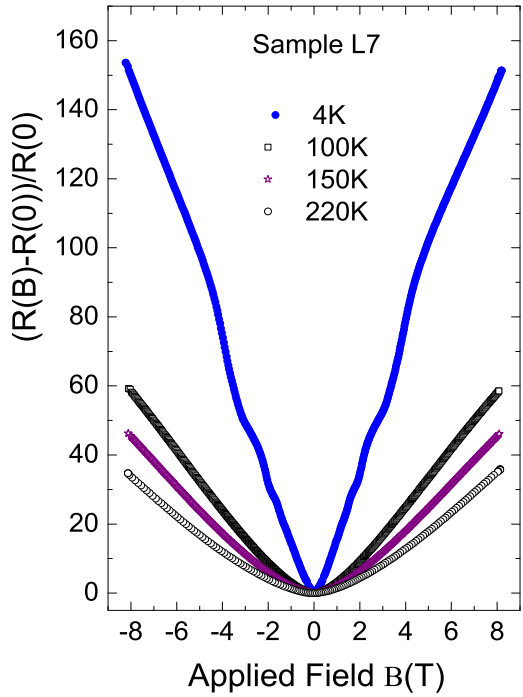}
\caption{The magnetoresistance  at different constant temperatures
for sample L7. The selected voltage electrodes were those of the d3
configuration, see inset in Fig.~\protect\ref{OMRL7size}.}
\label{omrl7}
\end{figure}

A qualitative and quantitative different behavior of the MR is
obtained for the metallic-like sample L7, see Fig.~\ref{omrl7}. Here
the MR decreases with temperature and it is a factor of $\sim 50$
larger than in the thinner sample L5. Because both samples have
similar lateral sizes the difference in behavior in the MR should be
related to the difference in thickness, i.e. in the number of
interface regions within the sample that also may influence the
resistivity. We show below that the MR can be related to the
resistivity of each sample.

\begin{figure}[htb]%
\includegraphics*[width=\linewidth,height=7.5cm]{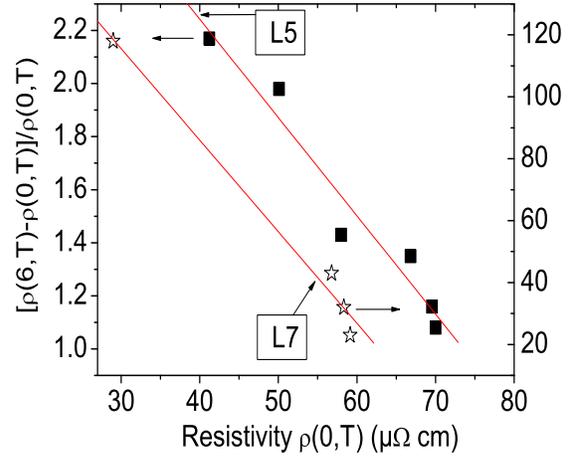}
\caption{The magnetoresistance at 6~T field vs. the resistivity at
zero field and at different constant temperatures for samples L5 and
L7 (right $y-$axis).} \label{deltarho}
\end{figure}

Figure~\ref{deltarho} shows the MR for both samples L5 and L7 at 6~T
field at different temperatures as a function of the resistivity at
the same temperatures. This figure shows that the decrease or
increase of the MR with $T$ is related to the
 resistivity value at zero field. This result appears to be compatible with the
semi classical picture for the magnetoresistance, i.e. the longer the
relaxation time -- or the smaller the resistivity -- the larger can
be the effect of the field on the resistance.

As shown in Fig.~\ref{deltarho} the MR of multigraphene samples as
well as in HOPG bulk samples \cite{kempa02} does not follow the well
known Kohler's rule. This rule describes the MR as a functional
scaling with the ratio $B/\rho$, i.e. in lowest order the MR $\propto
(B/\rho(0))^2$. There are at least three different reasons for the
failing of the Kohler rule in graphite, namely:\\
- One reason is related to the huge electron mean free path $\ell(T)$
in graphite, which is much larger than the radius of curvature of an
electron orbit under a magnetic field $r_c = m*v_F/eB$, with $m*
\lesssim 0.01 m$ the effective electron mass, $v_F \sim 10^6~$m/s the
Fermi velocity and $e$ the electron charge. For a field of 1~T, for
example, we have $r_c \sim 0.1~\mu$m whereas $\ell > 0.1~\mu$m for $T
< 200~$K at zero field \cite{gar08}.\\
- Other reason is that the semiclassical picture of the MR breaks
down when the Fermi wavelength $\lambda_F$ gets larger than $r_c$,
i.e. what is the meaning of the classical cyclotron radius $r_c \sim
0.1~\mu$m at $B \sim 1~$T when the wavelength of the electrons
$\lambda_F \gtrsim 0.1~\mu$m below 200~K \cite{gar08}.\\
- A third effect might be related to the contribution of the sample
internal structure and interfaces. These defects may not only
influence the dimensionality of the transport in graphite (short
circuiting the graphene planes) but also might be the origin for
localized, granular superconducting regions. Recently done study of
the behavior of the MR of multigraphene samples suggests the
existence of granular superconductivity \cite{esqui08}. The main
experimental evidence comes from the anomalous irreversible behavior
of the MR, which appears compatible with Josephson-coupled
superconducting grains \cite{esqui08}.

We note that Bi, as graphite, has a low density, low effective mass
of carriers and huge values of the electron mean free path
\cite{fri67,gan68}. It shows a very similar magnetic field induced
metal-insulator transition and also has a very low resistivity
\cite{heb05}. We may speculate that a similar superconductivity
phenomenon may play a role. In fact recently published work found
that crystalline interfaces in Bi bicrystals of inclination type show
superconductivity up to 21~K \cite{mun06,mun08}. A similar situation
may occur at the interfaces between crystalline regions in oriented
graphite, see Fig.~\ref{TEM}.

An important point we would like to stress here is that the peculiar
behavior of the MR in HOPG samples \cite{yakovadv03} is observed only
for fields applied perpendicular to the graphene layers. For fields
parallel to the graphene layers there is no MR, i.e. the measured
very weak MR can be explained by the misalignment of the field with
respect to the graphene planes \cite{kempa03}. This speaks for a huge
anisotropy of the assumed superconducting region. Therefore the
available experimental data suggest the interface regions as possible
candidates where this superconductivity might be located, see
Fig.~\ref{TEM}. Within the same schema it appears clear that the
decrease or even the level-off of the resistivity decreasing
temperature, see Fig.~\ref{rt}, would not be intrinsic but due to the
influence of the interface regions.

In case we have non-percolative superconducting arrays of grains, one
would expect to have a relatively larger increase in the resistance
with magnetic field mostly in the temperature region where the
influence of the coupling between superconducting grains at no
applied field is observable, i.e. in a region where either the
resistance decreases decreasing temperature or it levels off below a
certain temperature as for samples L5 and L8A below 25~K or below
30~K for sample L2A, see Fig.~\ref{rt}. This is expected for granular
superconductors due to a coherent charge transfer of fluctuating
Cooper pairs between the grains \cite{ler08}. This sensitive change
of the temperature dependence of the resistance under an applied
magnetic field has been already seen in Ref.~\cite{yakov99} for bulk
HOPG samples. The possibility of superconductivity in graphite has
been discussed in recent years, see Refs.~\cite{kopejltp07,esqui08}
for further reading.

\begin{figure}[htb]%
\includegraphics*[width=\linewidth,height=8cm]{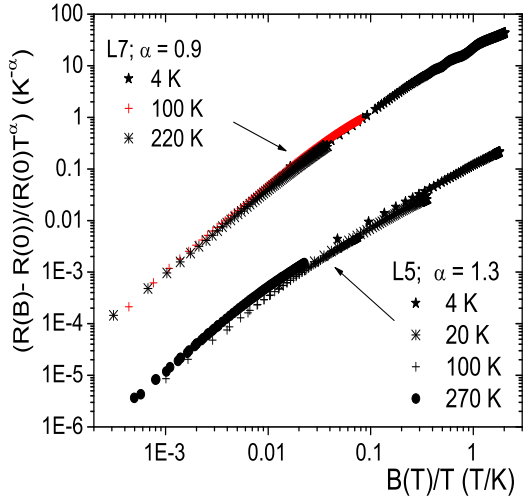}
\caption{The magnetoresistance divided by the temperature to the
exponent $\alpha$ vs. the ratio of the applied field to  temperature
$B/T$ for samples L5 and L7. The scaling is achieved for both samples
using two different exponents $\alpha = 0.9 (1.3)$ for sample L7
(L5).} \label{scal}
\end{figure}

On the other hand, in a multigraphene sample we certainly have other
defects that would not trigger local superconductivity but increase
the carrier density, decreasing $\lambda_F$ and therefore for a large
enough sample (see section \ref{lat}) the MR should increase.
Simultaneously, reducing $\lambda_F$ the Schubnikov-de Haas (SdH)
oscillations in the MR should be recovered, as seen in sample L7 at
low temperatures and high fields, see Fig.~\ref{omrl7}. This is
similar to the effect observed in graphene layers by applying a large
enough bias voltage, increasing the carrier density, decreasing
$\lambda_F$ and recovering the SdH oscillations \cite{novo05}.

Finally, we would like to remark an interesting aspect of the MR
related to its field and temperature dependence.  The field
dependence of the MR is quasi-linear in field at low enough
temperatures and high enough fields $(B > 0.01~$T), see for example
the MR for sample L5 in Fig.~\ref{omrl5}. This behavior is still not
well understood; possible explanations are based on Landau level
quantization of the Dirac fermions \cite{abri00} or due to the
circulations of quantum, inhomogeneous current paths at the borders
of graphite platelets creating a Hall-like voltage contribution to
the MR \cite{kempa06}. As mentioned above, neither the MR of the
multigraphene samples studied in this work nor the one in bulk
graphite follows the classical Kohler scaling. We have found however,
that the MR data for the multigraphene samples L5 and L7 show an
impressive scaling when $(R(B)-R(0)/(R(0)T^\alpha)$ is plotted vs.
the ratio $B/T$, as can be seen in Fig.~\ref{scal} for $\alpha = 1.3$
and $0.9$, respectively.

This kind of universal scaling of the magnetoresistance  has been
observed previously in different materials near a metal-insulator
transition, like in metallic Si:B ($\alpha = 1/2$) \cite{bog95},
icosahedral AlPdRe ($\alpha \simeq 0.3$) \cite{sri02} or in
intercalated amorphous carbon ($\alpha = 1/2$) \cite{kum06}. Whereas
the scaling with $\alpha = 1/2$ has a physical interpretation
 based on electron-electron interactions \cite{bog95}, deviations
 from this value remain unexplained. The scaling obtained for samples L5 and L7
is indeed extraordinary since it covers several orders of magnitude
in both scaled axes and appears to be unique in the literature.
Future studies should clarify whether this scaling is related to the
influence of superconductivity in the MR.

\subsection{Lateral Size dependence of the
Magnetoresistance}\label{lat}

As has been shown in recent work on HOPG bulk samples\cite{gon07} the
lateral size of the sample has an influence on the MR. This can be
qualitatively understood if we take into account that: (a)  The
de-Broglie wavelength for massless Dirac fermions $\lambda_D \simeq h
v_F/E_F$ ($v_F \simeq 10^6~$m/s is the Fermi velocity and a typical
Fermi energy $k_B E_F \lesssim 100$~K) or for massive carriers with
effective mass $m^\star \lesssim 0.01 m$, $\lambda_m =
h/\sqrt{2m^\star E_F}$, as well as the Fermi wavelength $\lambda_F
\sim (2\pi/n)^{1/2}$ can be of the order of microns or larger due to
the low density of Dirac and massive fermions. (b) Due to the
extraordinary large values of the mean free path of the carriers in
bulk HOPG ($0.4^\circ$) $\ell [\mu$m$] \sim ((10~\mu$m$)^{-1} + (6
\times 10^3 T^{-2})^{-1})^{-1}$ \cite{gar08}, at low enough
temperatures we expect to see a decrease of the MR reducing the
lateral sample size or the distance between the voltage electrodes.

\begin{figure}[htb]%
\includegraphics*[width=\linewidth,height=10cm]{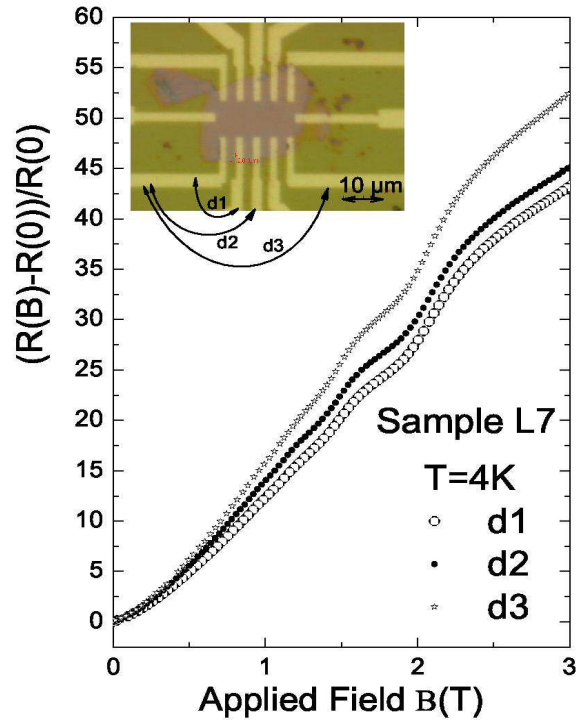}
\caption{The magnetoresistance of sample L7 at 4~K vs. applied field,
obtained at three different voltage pairs (d1 to d3) as shown in the
optical microscope picture in the inset.} \label{OMRL7size}
\end{figure}

Figure~\ref{OMRL7size} shows the MR of sample L7 measured at three
different voltage electrode positions d1, d2 and d3, see inset. As
expected, the larger the distance between voltage electrodes the
larger the MR. Note that a clear effect is observed changing the
distance from $\sim 4$ to 16~$\mu$m indicating that the mean free
path should be of this order, in agreement with recently published
results \cite{gar08}.

\begin{figure}[htb]%
\includegraphics*[width=\linewidth,height=10cm]{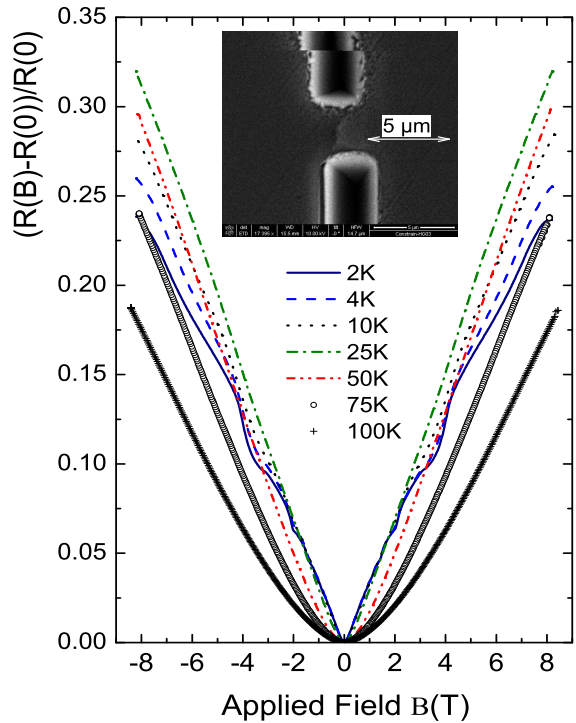}
\caption{Magnetoresistance  for a $17 \pm 2~\mu$m thick HOPG sample
at different temperatures with a constriction. In the middle of the
HOPG sample and between the voltage and current electrodes a
constriction was prepared by focussed Ga$^+$-ion beam. The inset
shows a scanning electron microscope image of the constriction.}
\label{hopgcon}
\end{figure}

The lateral size effect on the MR is also nicely observed in bulk
HOPG samples with a constriction. Using the Ga$^+$-ion beam of the
DBM we have prepared a $\sim 3~\mu$m wide constriction at the middle
and between the sample electrodes, see inset in Fig.~\ref{hopgcon}
where the results for the MR at different temperatures are shown. The
MR of the HOPG sample with a constriction is about two orders of
magnitude smaller than in the original sample. Leaving by side the
influence of the SdH oscillations we note that the MR at $B \le 2~$T
remains nearly $T$-independent from 4~K to 25~K and decreases at
higher temperatures. We note also that at zero field the ratio
between the resistivities at 50~K and 4~K $\rho(4)/\rho(50) \sim
0.98$ for the sample with constriction, see Fig.~\ref{hopgcon},
whereas it is $\sim 0.3$ for the original sample, see Fig.~\ref{rt}.
Therefore, the behavior of the MR can be understood assuming that the
mean free path remains larger than the width $W$ of the constriction
at $T < 25~$K and is smaller at higher $T$. This lateral size effect
on the MR indicates that the effective mean free path of the carriers
responsible for the MR should be of the order of $3~\mu$m at $\sim
25~$K, in excellent agreement with recently reported values for
similar HOPG samples \cite{gar08}. In case granular superconductivity
is confirmed in oriented graphite, future studies should clarify to
which extent this phenomenon contributes to the observed lateral size
effect in the MR.

\section{Conclusion}\label{con}

The data obtained in this investigation imply that multigraphene
samples show two size effects. The smaller the thickness of the
multigraphene sample the larger is the resistivity. The correlation
between the thickness dependence of the resistivity and the
microstructure of highly oriented pyrolytic graphite suggests that
the interfaces between crystalline regions and parallel to the
graphene layers could be the regions where granular superconductivity
is located. The differences of the temperature as well as the
magnetic field dependence of the resistivity of different
multigraphene samples suggest that those interfaces play a main role.
The available data indicate that the intrinsic $T$-dependence of the
resistivity of ideal multigraphene or graphite would be
semiconducting-like. This would imply that interpretations based on
the metal-like transport properties of graphite as well as
multigraphene samples should not be related to the properties of an
ideal graphite structure. The lateral size dependence of the
transport properties, specially in the magnetoresistance, can be
understood taking into account the large effective mean free path of
the carriers.

\begin{acknowledgement}
It is a pleasure to thank Dr. S. Reyntjens from the FEI company in
Eindhoven for providing us with the TEM images of the HOPG sample.
Fruitful discussions with N. Garc\'ia are gratefully acknowledge. We
gratefully thank  U.~Teschner and W.~Grill for the Raman
measurements. This work was supported by the DFG under DFG ES
86/16-1. J.-L. Yao acknowledges the support of the A. v. H.
Foundation and J. B-Q. the support of the EU project ``Ferrocarbon".
\end{acknowledgement}

%
%

\providecommand{\WileyBibTextsc}{}
\let\textsc\WileyBibTextsc
\providecommand{\othercit}{} \providecommand{\jr}[1]{#1}
\providecommand{\etal}{~et~al.}

\end{document}